\documentclass[%
 aip,
 amsmath,amssymb,
 reprint,%
]{revtex4-1}

\usepackage{graphicx}
\usepackage{dcolumn}
\usepackage{bm}

\usepackage[utf8]{inputenc}
\usepackage[T1]{fontenc}
\usepackage{mathptmx}
\usepackage{subcaption}
\usepackage{PLJ}
\usepackage{hyperref}

\begin{document}

\preprint{AIP/123-QED}

\title[WPDR Topology at 2nd harmonic resonance]{Topology of the Warm plasma dispersion relation at the \snd Harmonic Electron Cyclotron Resonance Layer}

\author{P.L. Joostens}
\affiliation{
Faculty of Applied Physics, Eindhoven University of Technology (TU/e), De Zaale, 6300 MB Eindhoven, Netherlands}
\author{E. Westerhof}
 \email{E.Westerhof@differ.nl}
\affiliation{
DIFFER - Dutch Institute for Fundamental Energy Research, De Zaale 20, 5612 AJ Eindhoven, the Netherlands}

\date{\today}

\begin{abstract}
The Warm Plasma Dispersion Relation, for waves in the electron cyclotron resonance range of frequencies, can be cast into the form of a bi-quadratic equation for $N_\perp$, where the coefficients are a function of $\nps{}$ and an iterative procedure is required to obtain a solution. However, this iterative procedure is not well understood and fails to converge towards a solution at the \snd harmonic resonance layer. In particular at higher densities where the wave can couple to an electron Bernstein wave.
This paper focuses on a solution to the poor convergence of the iterative method, enabling determination of the topology of the dispersion relation around the \snd harmonic using a fully relativistic code for oblique waves.
A feed-forward controller is proposed with the ability to adjust the rotation of a step of $\nps{}$ within the complex plane, while also limiting the step-size.
It is shown that implementation of the controller stabilizes unstable solutions, while improving overall robustness of the iteration. This allows the evaluation of the coupling between the fast extraordinary mode and electron Bernstein waves at the \snd harmonic electron cyclotron resonance layer, for non-perpendicularly propagating waves.
\end{abstract}

\maketitle

\section{\label{sec:Intro}Introduction}

Many codes exist to evaluate the dispersion of plasma electron cyclotron (EC) waves, all making use of their unique set of physics-based models and underlying assumptions \cite{Figini-2012,Prater-2008}. It has been shown that the cold plasma approximation is sufficient when describing wave propagation for ITER relevant scenarios, in which O-mode wave injection at the fundamental harmonic (O1) EC resonance is used \cite{Prater-2008}. Calculation of the wave power deposition, however, always requires the solution of the warm plasma dispersion relation. There are many devices that heat at the \snd harmonic X-mode (X2) because of the inaccessibility of the O1 resonance due to the plasma being critically dense. Also large future reactors might benefit from X-mode wave injection at large angles near the X2 resonance for the purpose of electron cyclotron current drive, provided gyrotrons capable of generating such high frequencies are available in the future. Evaluation of prolonged wave propagation near a resonant structure requires warm plasma effects to be taken into account, by means of assuming a particle velocity distribution leading to a warm plasma dispersion relation (WDPR). Between the \snd harmonic and the right-handed cutoff warm plasma effects give rise to the presence of Electron Bernstein Wave modes (EBWs) coupling to the fast X-mode (FX), creating complex structures within the solution space. Mapping the topology of the FX- and EBW-dispersion surfaces is therefore required to increase the understanding of these structures allowing for the validation and benchmarking between different codes \cite{Prater-2008,Prater-2004}. Furthermore a completion of the mapping  of the WPDR would provide a theoretical base, that allows for its  substitution with a neural network that resembles the behavior of the WPDR, while decreasing calculation time. Such a reduction in calculation time would make warm-plasma ray-tracing suitable for real-time applications \cite{Poli-2018}.

The work done for (close to) perpendicular propagating waves revealed most of the features of the dispersion relation's topology. Such as the work of Imre and Weitzner \cite{ImreWeitzner-1985a,ImreWeitzner-1985b} being the first to report on the crossing of the X-mode and the EBW-mode branches at the \snd harmonic, while also including a numerical evaluation of transmission reflection and absorption in the corresponding resonant region. It was later shown that the topology of the dispersion relation in terms of the complex valued perpendicular refractive index ($N_\perp$) could be evaluated using concepts from complex function analysis, where it became evident that the FX and the EBW solutions branches both formed a Riemann-like surface \cite{Egedal-1994}. Those surfaces could recombine to form an avoided crossing thus coupling the FX-mode and the EBW-mode at the second harmonic, which we will refer to as the FX2-EBW connection. When coupled the low field side (lfs) FX-mode is attached to the high field side EBW (the lfs-FX-EBW branch) and the high field side (hfs) FX-branch is coupled to the lfs-EBW branch (the hfs-FX-EBW branch). Although this analysis provided a good understanding for purely perpendicular propagating waves, further assessment of the FX2-EBW connection still needs to be done for waves under an oblique angle with respect to the toroidal magnetic field. Since EBWs are longitudinal waves that are generated by electrons coherently gyrating around their guiding centre it is expected that for some oblique wave angle the coupling between the FX-mode and EBWs is lost.

The structure of this paper is as follows. In section 2 a brief description of EC wave dispersion is given, that focuses on the way roots to the dispersion relation can be found. As will turn out, the presented root-finding method not always converges to a solution, hence requires an improvement in the form of a feed-forward controller. This improvement is explained in Section 3. Section 4 addresses the topology of the WPDR in terms of a normalized magnetic field and normalized density, similar to \cite{Egedal-1994}, focusing on the relation between the FX2-EBW coupling with plasma density, temperature and parallel refractive index.  Lastly Section 5 elaborates on the optical thickness of the plasma in the X2 resonant domain, to discuss the relevance of a proper working root-finding method for accurate power deposition calculations in ray-tracing codes, followed by a summary and conclusions.

\section{Electron cyclotron wave dispersion}

\subsection{Fully relativistic case}
Mapping of the WDPR is done by using the warm plasma routines from the 3D TORAY-FOM ray trace code \cite{Kritz-1982,Westerhof-1989}. In TORAY-FOM beam propagation is evaluated using a set of multiple rays, where each individual ray is evaluated according to geometrical optics, using the perpendicular refractive index $N_\perp$ determined by evaluating the WPDR. For evaluation of the dispersion of electromagnetic waves in the EC range we can ignore the negligibly small ion contributions to arrive at the general dispersion relation:
\begin{equation}
    \label{eq:general_dispersion}
    \hspace{8px} D \equiv |\varepsilon + N^2 ( \hat k \hat k - I)| = 0.
\end{equation}
\noindent The cold approximation of the dielectric tensor $\varepsilon$ -  where temperature effects to the dispersion of the EC wave are neglected - allows one to find two well known modes that are distinguished by the orientation of the electric field component of the injected wave ($\overrightarrow{E_1}$) w.r.t. the magnetic field direction inside the plasma ($\overrightarrow{B_0}$). The ordinary mode (O-mode) where $E_1 \parallel B_0$ and the extraordinary mode (X-mode) where $E_1 \perp B_0$. With the dielectric tensor $\varepsilon$ chosen to take warm plasma effects into account, an infinite number of solutions exist in terms of $N_\perp$ for wave propagation. Apart from the ordinary an extraordinary modes, the remainder of the solutions are known as Electron Bernstein Waves and have initially been discovered in 1958 by I.B.~Bernstein \cite{Bernstein-1958}.

To find both of X-mode and EBW solutions we use the fully relativistic dielectric tensor as published by Farina, in which the dielectric tensor is constructed in a manner that allows for evaluation at any order for the Bessel function expansion of the finite Larmor radius effects \cite{Farina-2008}. In the present work we have set the finite Larmor radius expansion to 5 which is more than enough to account for warm plasma effects around the \snd harmonic resonance, for temperatures up to a few 10's of keV as relevant for tokamak fusion reactors.

The dispersion relation can be regarded as a function of the from D($N_\perp^2$,$N_\parallel^2$,$T_e$,$X$,$Y$), where $ X = (\frac{\omega_{pe}}{\omega})^2 $ and $Y = \frac{\omega_{ce}}{\omega}$ represent the normalized plasma (electron) density and normalized magnetic field strength respectively. The X,Y-plane forms the Clemmow-Mullaly-Allis (CMA) parameter space which we  will refer to as the CMA-plane throughout this paper. The remainder of parameters and further definitions are as follows: electron temperature $T_e$, refractive index $N \equiv kc/\omega$, perpendicular refractive index $N_\perp$, parallel refractive index $N_\parallel$, wave vector $k$, unity wave vector $\hat{k}$, angular wave frequency $\omega$, electron plasma frequency $\omega_{pe} \equiv \sqrt{n_e q_e^2 / m_e \epsilon_0}$, electron cyclotron frequency $\omega_{ce} = |q_e| B / m_e$, magnetic field $B$, electron density $n_e$, electron mass $m_e$, electron charge $q_e$, and permittivity of vacuum $\epsilon_0$.

\subsection{Finding roots to the dispersion relation} \label{sub:dispersion roots}

The dispersion relation as presented in Equation \ref{eq:general_dispersion} can be rewritten into a bi-quadratic equation to find roots of $\nps{}$ as a function of ($N_\parallel^2$,$T_e$,$X$,$Y$):
\begin{equation}
    \label{eq:biquadratic}
    \hspace{8px} A(\nps{} ) N_\perp^4 + B ( \nps{} ) \nps{} + C(\nps{}) = 0.
\end{equation}
\noindent Detailed expressions for the coefficients are given in the appendix and are chosen such that around the fundamental harmonic and taking into account only the lowest significant order in the Larmor radius expansion of the Bessel functions, the coefficents are fully independent of $\nps{}$ \cite{fidone_1978,Westerhof-1989}. The inclusion of higher harmonic resonances and the multi order expansion of the Bessel function results in the coefficients A, B and C themselves being functions of $N_\perp^2$. This requires an iterative procedure for solving the bi-quadratic equation. An initial $\nps{,0}$ is assumed to calculate the complex coefficients A, B, and C to arrive at two roots for $N_{\perp,1}^2$ \cite{Matsuda-1991,Farina-2008}. One of which corresponds to O-mode propagation and the other to X-mode propagation and EBWs combined. To properly distinguish solutions between these modes the discriminant ($B_i^2 - 4 A_i C_i$) is checked whether it alters between quadrant 2 and 3 within the complex plane as an indication that the sign describing the desired mode has to be changed in the numerical evaluation of the square root function, in order to remain on the proper Riemann surface. The newly found value is then used for the recalculation of the coefficients in an iterative manner:
\begin{equation}
    \label{eq:biquadratic2}
    \hspace{8px} \nps{,i+1} = \frac{ -B_i \pm  \sqrt{B_i^2 - 4 A_i C_i} }{ 2 A_i  },
\end{equation}
\noindent where the implicit coefficients are defined as:
\begin{equation}
    \label{eq:coefficients}
      \hspace{8px} Z_i \equiv Z(\nps{,i}) \hspace{1cm} Z = A,B,C.
\end{equation}
\noindent The solution for $\nps{}$ is said to be converged to a solution when:
\begin{equation}
    \label{eq:tolerance}
    \hspace{8px} |\nps{,i+1} - \nps{,i}| < {\rm tolerance}.
\end{equation}
\noindent This method of iteratively solving the bi-quadratic from of the dispersion function is currently used not only used in the 3D multi-ray TORAY-FOM code, but the method also resides in the quasi-optical beam tracing code GRAY \cite{Farina-2005} and the real-time capable TORBEAM code \cite{Poli-2001,Poli-2018}. In the case of an undamped wave the iterative method quickly converges towards a solution. This convergence towards a solution however, does not rest on a theoretical basis and should not be considered trivial. In fact, near the second harmonic resonance, where the imaginary part of the square of the perpendicular refractive index is larger than zero, Im$\{ \nps{} \} >  0$, conditions for finding a solution become more stringent as the FX- and EBW-solutions approach each other. It is in this region that the iterative method fails to converge. In order to allow for a complete mapping of the WPDR and accurate warm plasma ray tracing around the \snd harmonic electron cyclotron resonance, modifications to the iterative method are required that ensure convergence towards the solution-branch as followed along a wave trajectory.

\section{Improving the iterative method}\label{sec:StabilityOf} \label{sec:ImprovingIterativeMethod}


The inability of the iterative method to converge towards a solution is most profound around the second harmonic electron cyclotron resonance structure, where the imaginary part of $\nps{}$ becomes significant compared to its real part. Especially at plasma densities low enough for low-field side (lfs) X-mode injection to reach the \snd harmonic electron resonance but high enough for the FX2-EBW coupling to exist, convergence to the desired solution does not take place. In fact an overshoot of the solution creates oscillating values of $\nps{}$ during iteration escaping the local domain of the desired solution and ending up at a different, undesired solution-branch to the bi-quadratic equation. This Section evaluates the iterative method as a dynamical system addressing the stability of solutions in Section \ref{sub:Stability}, where subsequently a controller is proposed in Section \ref{sub:Controller} that influences the iterative method by means of altering the step-size by means of a parameter $\lambda$ and the step-direction within the complex plane of $\nps{}$ through a rotation angle $\xi$. The proper setting of these parameters for a robust convergence of the iterative method is further detailed in Section \ref{sub:Lambda} and Section \ref{sub:Xi}, respectively.

\subsection{stability} \label{sub:Stability}
The trajectory traced out by the subsequent steps in the iterative solution of the bi-quadratic equation can be regarded as a trajectory of a dynamical system, thus allowing to apply the mathematical concepts of stability of dynamical systems. An approach is to evaluate solutions and their surrounding topology by means of their gradient ($\nps{,i}-\nps{,i-1}$) in the complex plane. A solution can then be classified as stable (unstable) when the gradient converges (diverges) around the solution.

\begin{figure*}
    \centering
    \includegraphics[width = 0.8\textwidth]{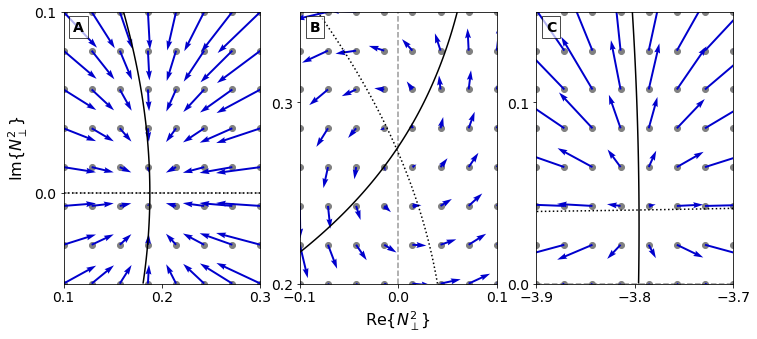}
    \caption{Solutions along the lfs FX-branch (X = 0.4, $N_\parallel$ = 0,$T_e$ = 3 keV), being coupled to the EBW-branch going to a cutoff showing: \textbf{A)} good convergence for small $\mathrm{Im}\{N_\perp \}$ (Y = 0.5), \textbf{B)} marginal stability of the solution for $\mathrm{Im}\{N_\perp \} \approx \mathrm{Re}\{N_\perp \}$ (Y = 0.512) and \textbf{C)} instability of the solution for $\mathrm{Im}\{N_\perp \} >> \mathrm{Re}\{N_\perp \}$ (Y = 0.540). Solutions of the dispersion relation are located at the crossing of the $\mathrm{Re}\{D\} = 0$ (black solid) $\mathrm{Im}\{D\} = 0$ (black dotted) contour lines. The blue arrows indicate the direction and step size of the iteration, starting from each of the grey dots.}
    \label{fig:1unstable}
\end{figure*}

To understand the lack of convergence for the lfs FX-mode branch, three distinct cases are plotted in figure \ref{fig:1unstable} for a constant normalized density of X = 0.4 and pure perpendicular wave propagation $N_\parallel = 0$ at an electron temperature $T_e$ of 3 keV. Solutions to the dispersion relation are found by imaging separately the zero crossings of the real and imaginary parts of D in the complex $\nps{}$ plane. Since a proper solution to the dispersion relation requires both real and imaginary parts of D to be equal to zero, solutions are identified by the intersection of the contours of $\mathrm{Re}\{D\} = 0$ (black solid line) and $\mathrm{Im}\{D\} = 0$ (black dotted line). The gradients of the iterative method are visualized using a quiver plot, where the quivers indicate the speed and direction of convergence.

The sub-figures in \ref{fig:1unstable}, A, B and C correspond to a normalized magnetic field Y of 0.5, 0.512 and 0.54 respectively and are chosen such that in A the imaginary part of $N_\perp$ equals zero and that in B the imaginary and real part of $\nps{}$ are almost equal to each other. In C $\mathrm{Re}\{\nps{}  \}$ is largely negative while $\mathrm{Im}\{\nps{}\}$ is small, thus - after taking the root - leaves us with a solution to the dispersion relation in terms of $N_\perp$ that has an imaginary part that is larger than its real part.

It becomes evident that for the undamped wave at figure \ref{fig:1unstable}.A we have our solution in the centre of a converging vector field, hence our solution is stable and initial values for $\nps{}$ within the local vicinity of this solution converge efficiently towards the solution. In the case of figure \ref{fig:1unstable}.B the solution is marginally stable: close-by evaluations of the solutions gyrate around the solution without converging towards the centre. At figure \ref{fig:1unstable}.C the iterative method strongly diverges away from the solution. It should be noted that for the observed cases the gradient field near the solution is continuous, even for the negative domains of both the real and imaginary part of $\nps{}$. This enables the stabilisation of solutions for every initial guess for $\nps{}$ that lies in its local domain w.r.t. surrounding solutions.

\subsection{Adding a Controller} \label{sub:Controller}

In an effort to improve the convergence towards a local solution a controller was constructed that alters the step ($\nps{,c} - \nps{,i}$) taken in every iteration:
\begin{equation}
    \label{eq:biquadratic3}
    \nps{,c} = \frac{ -B_i \pm  \sqrt{B_i^2 - 4 A_i C_i} }{ 2 A_i  }.
\end{equation}
\noindent The length of the step is altered by a parameter $\lambda$ and its direction is rotated by angle $\xi$ according to equation:
\begin{equation}
    \label{eq:controller}
    \vnps{,i+1} = \vnps{,i} + \lambda \overrightarrow{\mathbf{R}}(\xi) (\vnps{,c} - \vnps{,i} ).
\end{equation}
\noindent Using a vector definition for the complex valued $\nps{,i}$:
\begin{equation}
    \label{eq:Ninvec}
   \vnps{,i} \equiv \begin{bmatrix}
           \mathrm{Re}\{ \nps{i} \} \\
           \mathrm{Im}\{ \nps{i} \} \\
         \end{bmatrix},
\end{equation}
\noindent and a 2D rotation matrix:
\begin{equation}
    \label{eq:Rmatrix}
    \overrightarrow{\mathbf{R}} \equiv \begin{bmatrix}
         \cos(\xi) && -\sin(\xi) \\
         \sin(\xi) && \cos(\xi) \\
    \end{bmatrix}.
\end{equation}
\noindent The definition for $\vnps{,i}$ in combination with the 2D rotation matrix $\overrightarrow{\mathbf{R}}$ leads to a rotation with angle $\xi$ in the complex plane. Note that $(\lambda,\xi) = (1,0)$ yields $\vnps{,i+1}=\vnps{,c}$ and thus, does not alter the iteration step with respect to the original iteration scheme (equation \ref{eq:biquadratic2}).

\begin{figure}[htb]
    \centering
    {\includegraphics[width=0.4\textwidth]{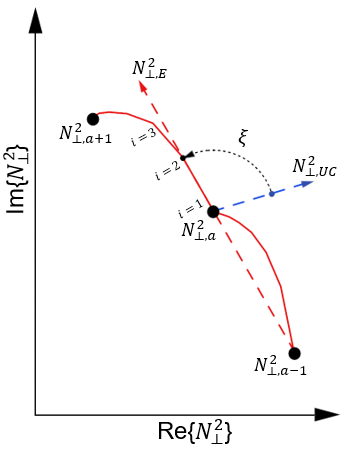}}
    \caption{Determination of the rotation angle $\xi$ as the angle between the initial uncontrolled step direction ($\nps{,UC}$ in blue) and the projection of the previous solution to obtain an expected solution ($\nps{,E}$ in red). Solutions to the dispersion relation along a ray trajectory (or a scan in Y) are marked with subscript a and the red line represents the trajectory between those solution using the controlled iterative method. $\xi$ allows us to find $\nps{,a+1}$, the next value along a ray, where the first iteration step starting from $\nps{,a}$ (marked as the small black dot) is located along the projection of the expected solution.}
    \label{fig:2XiAngle}
\end{figure}

To obtain a stable and robust iteration, a feedback loop is constructed that adjusts the parameters $\lambda$ and $\xi$ based on the corresponding local iterative topology of complex $\nps{}$. Since $\lambda$ dictates the step-size this generally influences only the speed of convergence, provided there is no overshoot leading to unstable behaviour. The rotation angle $\xi$ is determined as the angle between the direction in which a step tends to go and the direction the solution is expected to be. Although the direction from the initial location towards the solution is not known on forehand, a bordering solution to the bi-quadratic equation could be used as it is generally available when performing ray-tracing. A guess for the solution is obtained by linear extrapolation of two such previous solutions in the direction of the solution that is searched for. Figure \ref{fig:2XiAngle} illustrates this procedure when using the iterative method to converge to solution $N_{\perp,a+1}$, initiating from $N_{\perp,a}$ as obtained in the previous position along a ray-trajectory. From the solution at the earlier step $\nps{,a-1}$ a projection is made to estimate the direction where the next solution $N_{\perp,a+1}$ is expected labeled as $\nps{,E}$. Solving the dispersion relation using $N_{\perp,a}$ to calculate the coefficients $Z_a$ results in $\nps{,c}$. The direction of the iteration step ($N_{\perp,c} - N_{\perp,a}$) which needs to be rotated by $\xi$ to align with the direction where the solution is expected to be. This ensures that the first iterative step lies along the projection of the solution. Using this rotation angle at the following steps, the iteration now converges towards the solution as one would expect from Figure \ref{fig:1unstable}. To clarify, three steps are shown in the figure as i = 1,2 and 3 where i = 1 is the initial guess equal to the previous solution. Stabilisation of the iteration using the obtained $\xi$ results in the iterative method to convergence to the next solution $\nps{,a+1}$. After arriving at the new solution a new value of $\xi$ is obtained to calculate $\nps{,a+2}$ and so on. $\xi$ is changed after every solution and is kept constant while the iteration converges from one solution to the next.

\subsection{Decreasing step-size} \label{sub:Lambda}

Before evaluating the full controller the effect of merely adjusting the step-size controlled by $\lambda$ to smaller sizes is analyzed. Such a reduction in $\lambda$ reduces overshoot and extends the domain in Y in which the iterative method still converges towards the lfs-FX-EBW solution branch. To illustrate this, a point is taken within the X2 resonance relevant parameter space where FX-EBW branch coupling occurs, that lies close to the point where the FX- and EBW-mode connect ($X = 0.3$ at $3$ keV for $N_\parallel=0$). This ensures the lfs-FX and hfs-FX solution branches lie close together, creating challenging conditions in the topology of the step-directions within the complex plane of $\nps{}$.

\begin{figure}[htb]
    \centering
    \includegraphics[width = 0.5\textwidth]{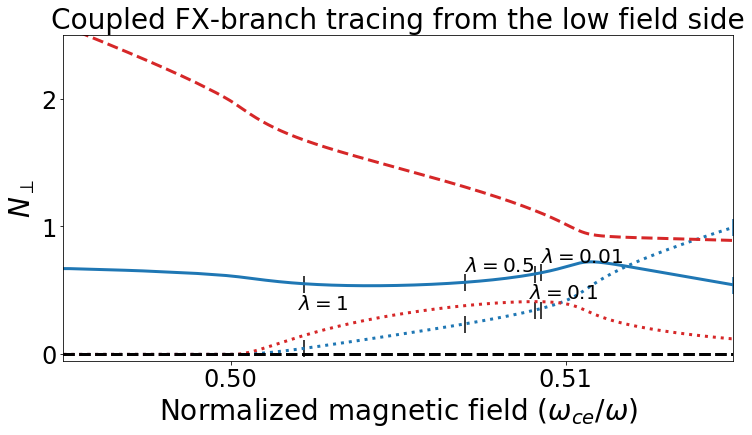}
    \caption{Lfs FX-EBW branch in blue (Re: \sampleline{}, Im: \sampleline{dotted}) obtained using different step-sizes $\lambda$, where smaller step sizes allow for a prolonged tracing along the branch up till the solution becomes unstable somewhere around Y = 0.51, hence a part of the branch cannot be found (blue dotted line) without controller. Hfs FX-EBW branch in red (Re: \sampleline{dashed}, Im: \sampleline{dotted}). ($X = 0.3$, $T_e = 3$ keV, $N_\parallel = 0$ and d$Y = 2 \; 10^{-5}$)}
    \label{fig:3lam_only}
\end{figure}

Results of the iterative method in this point of parameter space are shown in Figure \ref{fig:3lam_only} for different step-sizes $\lambda$ showing the corresponding points ($Y$,$N_\perp$) where the method fails to converge. It becomes evident that at the original step-size ($\lambda = 1$) convergence breaks far before the solution becomes unstable because evidently as the locations $\lambda = 0.5,0.1$ and $0.01$ suggests, a smaller step-size would allow the convergence of the solution to continue much further into the \snd harmonic resonance layer along the lfs FX-EBW branch. Finally due to the solution becoming (marginally) unstable - as previously shown in Figure \ref{fig:1unstable}.B - a decrease in step-size no longer suffices to extend the traceable part of the branch (blue dashed line). To continue branch tracing from this point onward a controller adjusting the rotation angle of the iteration step is essential.

\subsection{Rotation of the Step-direction} \label{sub:Xi}

The performance of the controller is evaluated in two ways. First, the topology of the iteration dynamics in the complex $\nps{}$ plane is shown, for multiple angles of $\xi$, which shows that initial values within the local region around a solution converge for a specific angle $\xi$. This allows us to confirm the ability that this solution can be made stable using the corresponding angle $\xi$. Secondly we investigate the performance of the methodology presented in figure \ref{fig:2XiAngle} by tracing solution branches around the X2 electron resonance from both the low- and high-field side.

\begin{figure*}[htb]
    \centering
    \includegraphics[width = 0.8\textwidth]{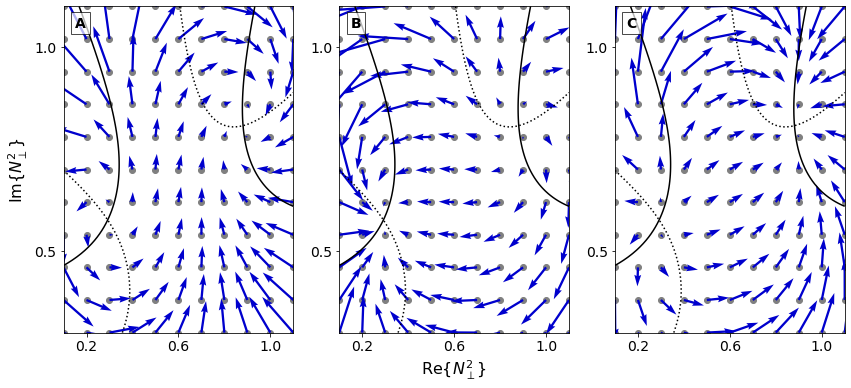}
    \caption{The lfs FX-EBW solution (located on the bottom left of each individual figure) and hfs FX-EBW solution (located on the top right of each individual figure) as determined from the crossing of the zero-crossing lines the real (full black curves) and imaginary (dashed black curves) pars of the bi-quadratic equation. Arrows in blue indicate the direction of a step of the iterative method starting from each gray dot, for corresponding stabilisation angles: \textbf{A)} ($\xi = 0^\circ$ no stabilisation) where the lfs-FX solution is unstable, the hfs-FX solution is stable and solutions from the vicinity of both solutions seem to end up at the hfs-FX solution. \textbf{B)} ($\xi = -98^\circ $) stabilizes the lfs-FX solution while rendering the hfs-FX solution unstable. \textbf{C)} ($\xi = 48^\circ$) improves the stability of the hfs-FX solution w.r.t. $\xi =0^\circ$. The two solutions used in this figure lie close together because $X =0.3$ and $Y = 0.51$ are close to the point where the FX-mode and EBW-mode coincide.}
    \label{fig:4stability}
\end{figure*}

To evaluate the effect of $\xi$ on a solution and its potential to stabilize it, the exemplary set of parameters that has been used earlier, to address the effect of $\lambda$ (Figure \ref{fig:3lam_only}) is used. On this branch around $Y = 0.51$ - where the original method fails to converge - the dynamics of the iterative method are examined. In Figure \ref{fig:4stability} this examination comprises of comparing the topology of the step-directions within the complex plane of $\nps{}$ (blue arrows) with the complex residue, which must be zero for a valid solution. Thus the zero-crossings of the real (full) and the imaginary (dotted) part of the residue are shown. An actual solution is represented as the crossing of these lines. In Figure \ref{fig:4stability}A no stabilisation is enabled ($\xi = 0$). The lfs FX-EBW solution (on the left of the complex plane) is unstable and iterations from points close by tend to converge away from it, possibly ending up at the hfs FX-EBW branch solution. The lfs-FX solution can however be stabilized by the introduction of a rotation of the step-direction, where an $\xi$ of -98$^\circ$ decreases gyration around the lfs-FX solution the most (Figure \ref{fig:4stability}B), while at the same time rendering the hfs-FX solution unstable. Furthermore convergence towards the hfs-FX can also be improved by stabilizing the gyration around this solution using a clockwise rotation $\xi = 48^\circ$ (Figure \ref{fig:4stability}C).

\begin{figure}[htb]
    \centering
    \includegraphics[width = 0.5\textwidth]{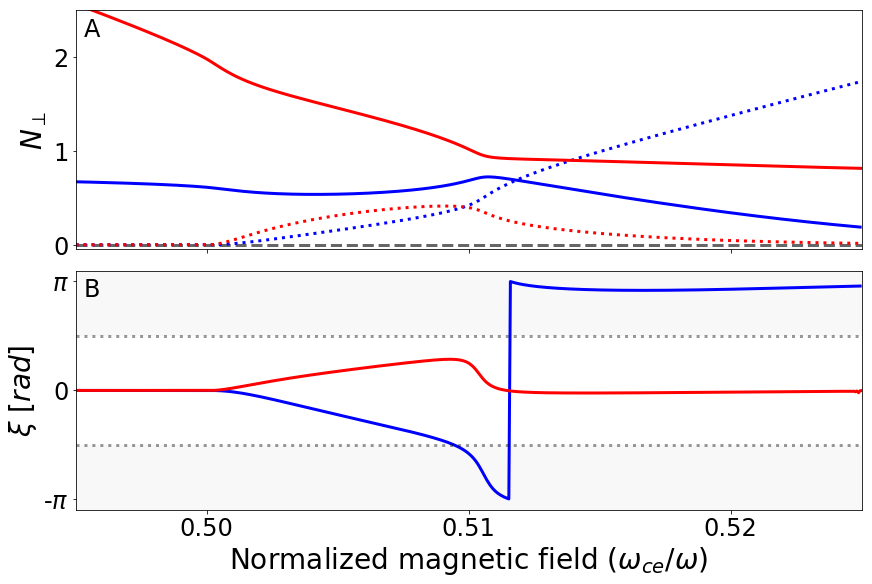}
    \caption{ \textbf{A:} Solution branches as obtained by the controlled iterative method.  \textbf{B:} Corresponding stabilisation angle $\xi$, where the gray area indicates instability of the solution ($|\xi + k 2\pi| > \pi/2, k \in \mathbb{Z}$). The lfs-branch (blue) goes to EBW cutoff after passing the second harmonic and the hfs-branch (red) connects to the EBW-mode.}
    \label{fig:5branchx3}
\end{figure}

Whereas Figure \ref{fig:4stability} indicates that a rotation of steps in the complex plane is able to stabilize a solution, branch tracing must be performed to confirm whether the feedback setup as in Figure \ref{fig:2XiAngle} stabilizes the entirety of a solution branch during ray tracing. For clarification, where branch tracing refers to the tracing of a solution branch of the dispersion relation while varying one or more input parameters ($P_{D}$) \{$X,Y,T_e,N_\parallel$\} $\in$ $P_{D}$, the term ray tracing refers to the following of a ray trough 6D (position, wave vector) \{$\overrightarrow{\mathbf{s}}$,$\overrightarrow{\mathbf{k}}$\} $\in P_R$ space where elements in $P_D$ are functions of $P_R$. So in order to evaluate the feasibility of the method used to stabilize the iterative method, branch tracing should suffice. Numerical performance with respect to ray tracing step size depends also on the relations between $P_D$ and $P_R$.
Branch tracing is performed over the normalized magnetic field $Y$, keeping $X$,$T_e$ and $N_\parallel$ constant. The most interesting branch for $N_\perp$ as a function of $Y$ is the EBW-connected lfs-FX branch as it has been shown that unstable solutions exist (Figure \ref{fig:1unstable}) and that those can be stabilized (figure \ref{fig:4stability}). The hfs FX-EBW branch, although not unstable, shows more robust convergence.

Figure \ref{fig:5branchx3}A shows both of these branches at $X = 0.3$ and $T = 3$ keV for perpendicular propagation obtained from the iterative method. Aside from the continuous behaviour of the branches an evaluation of the residue indicates that the controller - with stabilisation angle $\xi$ as shown in Figure \ref{fig:5branchx3}B - successfully stabilizes the solution. This is achieved in combination with a step-sizes of $\lambda$ = 0.1 and d$Y = 2 \; 10^{-4}$. Note how the point in Y where the solution becomes unstable ($|\xi| > \pi/2$ in gray) matches with the location of the loss of convergence for small step-sizes $\lambda$ that is shown in Figure \ref{fig:3lam_only}. Further inspection of the controller's state variable $\xi$ outside the \snd harmonic resonance layer indicates that a controller is only necessary for EBW waves. An Electron Bernstein wave that goes to cutoff requires a stabilisation angle of $180^\circ$ whereas EBWs propagating at high $N_\perp$ in between EC resonances can become marginally stable. Those regions are however not interesting from a practical point of view as these waves are strongly damped and cannot be accessed from outside the plasma. Lastly it should be stated that ordinary mode propagation does not benefit from a controller as described in this paper, as the iterative method already shows fast and robust convergence towards the O-mode solution.

\section{Extraordinary mode topology at the $2^{nd}$ harmonic resonance layer} \label{sec:Topology}

This section addresses the coupling of the FX-mode to the EBW-mode, which can be described using complex function analysis where both the FX-mode as well as the EBW-mode form a Riemann like surface within the CMA-plane. First the degenerate case of perpendicular wave propagation ($N_\parallel = 0$) is regarded in Section \ref{sub:Branches}, where - unlike previous studies - branch-cuts are placed parallel to the normalized magnetic field Y coordinate within the CMA-Plane. Then for a constant normalized plasma density X the coupling between the FX-mode and the EBW-mode is evaluated. Subsequently in Section \ref{sub:ObligueWaves} the domain in X where the modes are coupled will be mapped for oblique waves where $N_\parallel \neq 0$.

\subsection{FX-Branch connection to the EBW-mode} \label{sub:Branches}

To allow for the continuation of the work of mapping the WPDR and as an extension of the validation of the controller constructed in Section \ref{sec:ImprovingIterativeMethod} we zoom in at the points where the FX-mode and the EBW-mode coincide, as these 'branch' points form the boundaries between a detached FX-mode and EBW-mode and wave propagation where these modes are connected \cite{LazzaroRamponi-1981}. From preceding work using weakly relativistic codes and a lowest order approximation of the Bessel function for purely perpendicular propagating waves the dispersion tensor could be rewritten in the form of a bi-quadratic equation containing solely the X-mode solution and a single EBW-mode solution. This was convenient as the two roots to this bi-quadratic equation represent the FX-mode and EBWs separately. Hence coinciding FX- and EBW-branches could be found by solving $\mathrm{Re}\{\Delta\} = 0$ and $\mathrm{Im}\{\Delta\} = 0$, with $\Delta$ being the discriminant of the resulting bi-quadratic equation \cite{Bornatici-1981,Egedal-1994}. Although this method cannot be used for the fully relativistic dielectric tensor including higher order finite Larmor radius terms, nor for oblique waves, the controller designed in Section \ref{sec:StabilityOf} provides accurate solutions of our implicit bi-quadratic equation. From preceding work three branch points that are close to the R-cutoff meet the criteria for coincidence of FX- and EBW-mode. However only two will be evaluated in this paper as they fall within the \snd harmonic cyclotron resonance layer, whereas the third is associated with the first harmonic ($Y \approx 1$ for small $X$). These points are labeled $P_1 \equiv (X_1,Y_1)$ and $P_2 \equiv (X_2,Y_2)$, and are ordered such that $X_1 < X_2$ and $Y_1 < Y_2$. For the sake of clarity, within \cite{Egedal-1994} point $P_1$ was labeled $b_2$ and $P_2$ was labeled $b_3$.

\begin{figure*}[htb]

\includegraphics[width=0.4\linewidth]{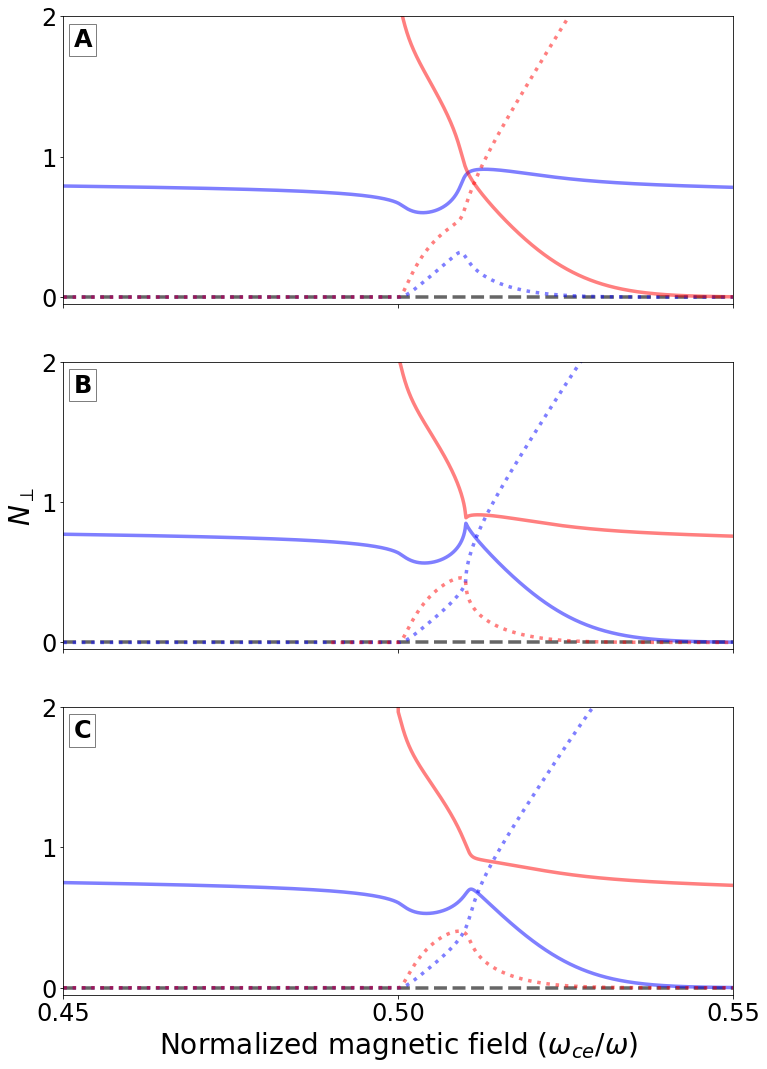}
\includegraphics[width=0.4\linewidth]{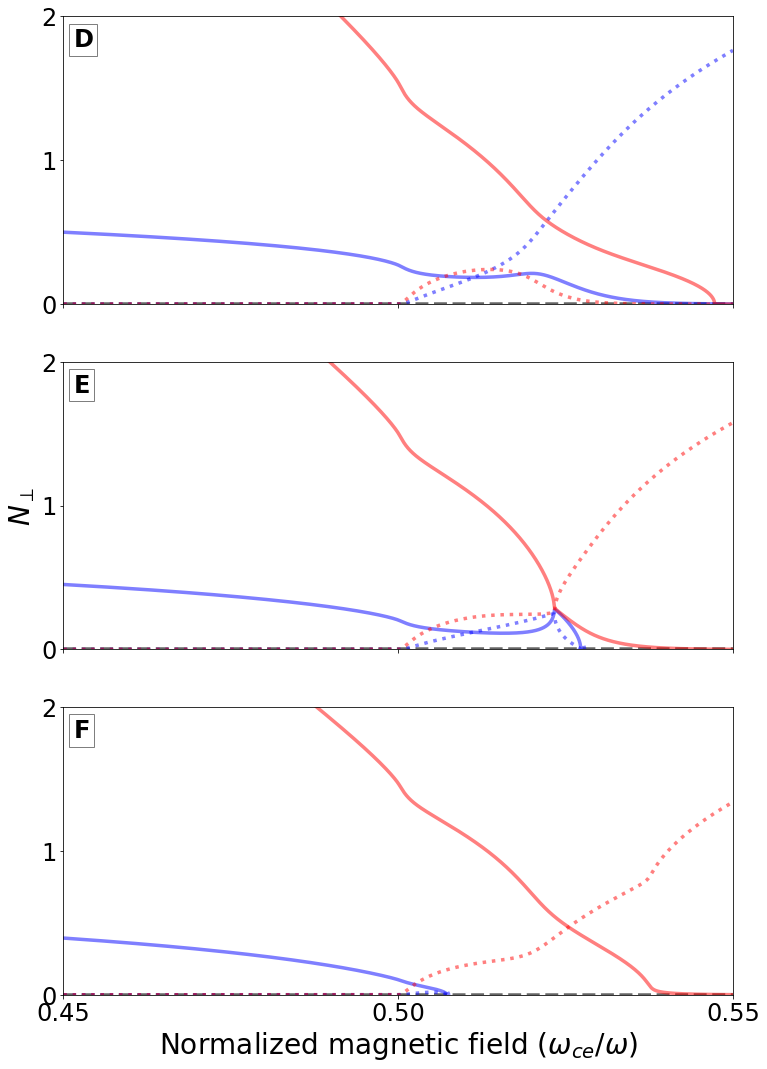}

\caption{Solution branches (\sampleline{}: $\mathrm{Re}$ \sampleline{dotted}: $\mathrm{Im}$) of the lfs-FX (blue) and the lfs-EBW (red) at: \textbf{A:} $X_1 - 0.02 \:$  \textbf{B:} $X_1 \:$  \textbf{C:} $X_1 + 0.02 \:$ \textbf{D:} $X_2 - 0.02 \:$  \textbf{E:} $X = X_2 \:$  \textbf{F:} $X = X_2 + 0.02 \:$}
\label{fig:F6_branches}
\end{figure*}

To convey the qualitative features of the FX2-EBW coupling, plots are made of $N_\perp$ as a function of $Y$ for constant normalized density $X$. Figure \ref{fig:F6_branches} contains six of these plots for the normalized density at $P_1$: $X_1 = 0.283$ (B) and close to $P_1$: $X = X_1 \pm 0.02$ (A,C) at a temperature of $3$~keV for a purely perpendicular propagating wave. The sub-figures D,E and F of \ref{fig:F6_branches} are equivalent to A, B and C but for the normalized density at $P_2$ where $X_2 = 0.487$. The qualitative features of the different branches $N_\perp(Y)|_{X,T_e,N_\parallel}$ are summarized below.
\begin{description}
    \item [Figure \ref{fig:F6_branches}A $X < X_1$] FX-mode waves are able to propagate past the X2 electron resonance layer from either side of the plasma. The propagation of the EBW mode is restricted to a narrow region around the \snd harmonic resonance layer, where the $\renp{}$ approaches zero and the mode becomes critically damped as it crosses the resonance layer.
    \item [Figure \ref{fig:F6_branches}B $X_1$] The modes coincide at $Y_1$, defining the point $P_1$.
    \item [Figure \ref{fig:F6_branches}C and \ref{fig:F6_branches}D $X_1 < X < X_2$] The crossing of the two wave branches is avoided by a coupling of the FX and EBW branches such that a FX wave propagating from the lfs connects to the critically damped part of the EBW branch, whereas the hfs-FX branch connects to the EBW branch going to high values for $N_\perp$ on the lfs of the \snd harmonic resonance layer.
    \item [Figure \ref{fig:F6_branches}E $X_2$] The modes coincide close to the R-cutoff at $Y_2$ such that the lfs- and hfs-FX branch reconnect again.
    \item [Figure \ref{fig:F6_branches}F $X > X_2$] The FX branch enters the R-cutoff before being able to connect to the EBW-mode.
\end{description}
Close to the branch points where the modes coincide, the gradient of both the real part and imaginary part of the perpendicular refractive index ($\frac{\partial N_\perp}{\partial Y}$) appears to go to infinity for FX- and EBW-waves inbound from the low field side. This gradient is related to the group velocity ($\frac{\partial\omega}{\partial k_\perp}$), where the group velocity tends to go to zero in these points:
\begin{equation}
    \label{eq:groupvel}
    \bigg|\frac{\partial \omega}{\partial k_\perp} \bigg| \propto \bigg|\frac{\partial N_\perp}{\partial Y}\bigg{|^{-1}}.
\end{equation}
\noindent This suggests that reflection of both waves will occur at these points. This is confirmed by the analysis of Imre and Weitzner, where it is shown that mode coupled reflection - where lfs-FX wave reflect along the lfs-EBW branch and vice versa - is highest for pure perpendicular wave propagation at low plasma temperatures and small length scales \cite{ImreWeitzner-1985b}.

\subsection{FX2-EBW connection for non-perpendicularly propagating waves} \label{sub:ObligueWaves}

For oblique waves where $N_\parallel \neq 0$ the FX-EBW connection vanishes once the wave angle w.r.t. $\mathbf{B_0}$ becomes too large. In Section \ref{sub:Branches} it is shown that the domain in which these two modes couple are bounded by the points $P_1$ and $P_2$. Analysis of sections of the FX- and EBW- branches over the normalized magnetic field is done by finding values for $X$, $T_e$ and $N_\parallel$ where the modes coincide, similar to figure \ref{fig:F6_branches}B and E. This allows the determination of $P_1$ and $P_2$ for nonzero $N_\parallel$ up till an order of $10^{-3}$ accurately. This accuracy was achieved in combination with a $dy$ and $N_\perp$ of the order $2 \; 10^{-5}$ and $10^{-4}$, respectively.

\begin{figure}[hbt!]
    \centering
    \includegraphics[width = 0.4\textwidth]{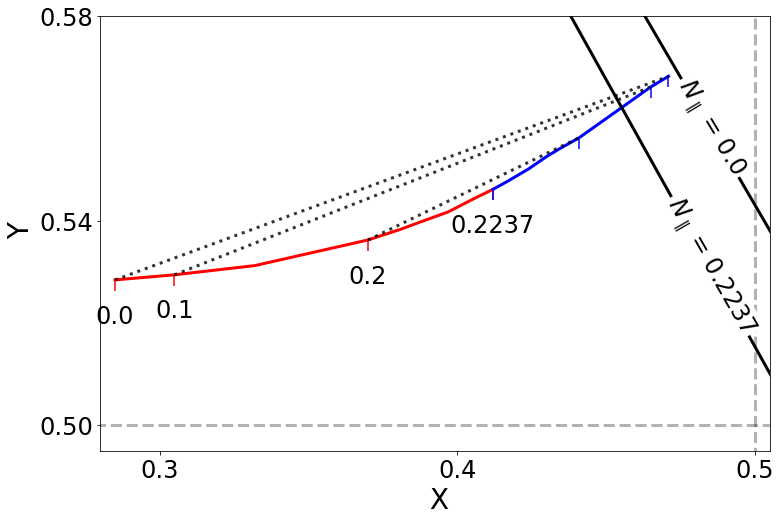}
    \caption{Locations of $P_1$ (red) and $P_2$ (blue) and the R-cutoff (black) in the CMA-plane as function of $N_\parallel$ at 9 keV. At $N_\parallel = 0.2237$ the branch-points merge and the FX2-EBW coupling vanishes. }
    \label{fig:F7_PointsInCMA}
\end{figure}

In Figure \ref{fig:F7_PointsInCMA} the locations of the branch points $P_1$ (red) and $P_2$ (blue) are shown in the CMA-plane as a function of $N_\parallel$ and for a constant temperature of 9~keV. For waves with an increasingly larger angle w.r.t. the normal of the magnetic field $\mathbf{B_0}$, $P_1$ and $P_2$ converge toward each other, creating more stringent conditions for the FX-EBW coupling. In the case of 9~keV the coupling ceases to exist at $N_\parallel = 0.2237$ as the branch-points merge together. Concurrently the R-cutoff (black line) shifts to a lower normalized magnetic field or equivalently a lower normalized plasma density for larger $N_\parallel$, in accordance with the cold plasma approach:
\begin{equation}
    \hspace{8pt} X_{cutoff} \propto (1-N_\parallel^2).
\end{equation}

\begin{figure*}[hbt!]
    \centering
    \includegraphics[width = 0.8\textwidth]{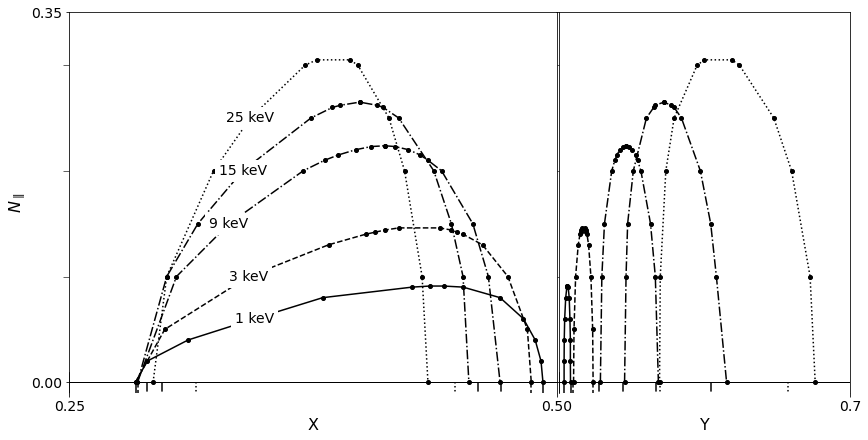}
    \caption{Locations of the branch-points for X- (left) and Y-coordinate (right) as a function of $N_\parallel$ for different plasma temperatures. Markings below the X-axis represent the $P_2$ fit from \cite{Egedal-1994} for the corresponding temperatures used in the figure. Dots in the figure represent the data-points used to construct the (dashed) lines by linear interpolation.}
    \label{fig:F8_PointsForT}
\end{figure*}

The exclusion of the FX2-EBW connection due to a parallel wave component was found to depend on the degree of relativistic broadening. In Figure \ref{fig:F8_PointsForT}  the locations of the branch points $P_1$ and $P_2$ are shown with their corresponding CMA coordinates X and Y in the left and right figure, respectively. The vertical axis represents the parallel refractive index. To illustrate trends in the data, lines in the graph are added as linear interpolation of data (dots) where the left side of the curves correspond to $P_1$ and the right side correspond to $P_2$ until they merge together in the centre of the curve. The area below the curves represents the domain ($X$,$Y$,$N_\parallel$) where the FX-EBW connection exists. It can be seen that at higher temperatures the FX2-EBW connection exists for increased oblique wave propagation (at 3~keV: $N_\parallel = 0.153$, at 9~keV: $N_\parallel = 0.227$ and at 15~keV: $N_\parallel = 0.266$). The profile of the Y coordinate shifts uniformly due to relativistic broadening, whereas the curve representing the relation between $X$ and $N_\parallel$ changes shape as the temperature decreases, since $X_1$ alters minimally while $X_2$ shifts along with the R-cutoff.

Along the horizontal line where $N_\parallel = 0$, the \snd order polynomial fit for $X_{1,2}$ and $Y_{1,2}$ as proposed in \cite{Egedal-1994} can be compared to the position of the points as evaluated for the fully relativistic case in the present work. Results from the polynomial fits are shown by markings below the X-axis in figure \ref{fig:F8_PointsForT} where the corresponding temperature is indicated by the same color. This reveals that the fit proposed does not match the fully relativistic data for temperatures above the warm plasma regime (e.g. $>$ 5~keV), as could be expected from their use of a first order estimate for the relativistic $\gamma$. However, what is interesting is that they predict a shift in $X_1$ for temperature whereas the fully relativistic approach indicates that there is no shift.

\section{X2 Wave absorption as determined by the iterative method} \label{sec:WaveAbsoptionIterative}

In this section the relevance of the improvements made on the iterative method for ray tracing and power deposition calculations is addressed. This is done by estimating the optical depth of the \snd harmonic resonance layer that is reached at the point where the original, unstable iterative method begins to fail for waves injected from the low-field side.

\subsection{Determination of the optical depth} \label{sub:OpticalDepthModel}


Transfer of energy between waves and plasma is generally described by an absorption coefficient $\alpha \propto Im\{N_\perp \}$ and the optical depth of a wave:
\begin{equation}
    \label{eq:tau1}
    \hspace{8pt} \tau = \int_{0}^{L(s)}  \alpha(s) \mathrm{d} s.
\end{equation}
Under the assumption that wave propagation occurs at constant electron density, the change in position (s) can be expressed as a function of Y using a typical length scale for the magnetic field gradient as:
\begin{equation}
    \label{eq:tau2}
    \hspace{8pt} L_B \equiv \frac{B_0}{|\partial B_0 / \partial s|}.
\end{equation}
Combined with an expression for the absorption coefficient $ \alpha(\mathrm{Re}\{\overrightarrow{\mathbf{k}}\}) = 2 \; \mathrm{Im} \{ k_\perp \}$ that holds for perpendicular propagating waves and $k_\perp = N_\perp \omega / c $ the optical depth $\tau(Y)$ can be expressed as a integral over Y:
\begin{equation}
    \label{eq:tau3}
    \hspace{8pt} \tau = 2\frac{\omega_{2}}{c} L_B \int_{0}^{L(Y)} \frac{\mathrm{Im}\{N_\perp\}(Y)}{Y} \partial Y.
\end{equation}
\noindent The absorbed power fraction $f_{P} \equiv P_{\rm absorbed}/P_{\rm wave}$ can be evaluated as:
\begin{equation}
    \label{eq:tau4}
    \hspace{8pt} f_{P} = 1-e^{-\tau}.
\end{equation}
The optical depth depends linearly on a tokamak's axial magnetic field strength $B_0$ where $\omega_2 [$rad/s$] = 2\pi \; 56 $[GhZ/T]$ B_0$ and  on a tokamak's specific major radius as $L_B \approx R_0$. The optical depth $\tau$ will therefore be normalized for the product of these parameters, including temperature $T_e$ as:
\begin{equation}
    \label{eq:tau5}
    \hspace{8pt} \tilde{\tau} = \frac{\tau}{L_B \; B_0 \; T_e}.
\end{equation}

\subsection{Failure of the iterative method and power deposition}

Quantification of the performance of the controlled and uncontrolled iterative method is established by means of locations of Y that are correctly evaluated for a particular method. This is convenient as for a waves originating from the low field side, the risk of overshoot of the solution becomes increasingly an issue as $\xi$ approaches $|\frac{1}{2}\pi|$, after which the solution becomes unstable (see figure \ref{fig:5branchx3}). Hence three locations in $Y$ for constant $X,T_e,N_\parallel$ are chosen following definitions as given below.
\begin{description}
    \item [$Y_a$] The uncontrolled iterative method fails to converge using the original step-size $\lambda = 1$.
    \item [$Y_b$] The uncontrolled iterative method fails to converge using a reduced step-size $\lambda = 0.1$.
    \item [$Y_c$] The controlled method finds a solution where $\mathrm{Im}\{N_\perp\} > \mathrm{Re}\{N_\perp\}$ and the wave becomes critically damped.
\end{description}
The corresponding optical densities at the locations labeled above are subsequently defined as $\tilde\tau_i$ for $Y_i$ where $i = a,b,c$, and are evaluated for the remainder of the parameters that apply to wave dispersion. In Section \ref{sub:Lambda} it has been explained that by reducing step-size  $\lambda$ solution within the stable domain of the branch can still be found. Therefore $Y_b$ and corresponding $\tilde\tau_b$ provides information whether a reduction of step-size could be sufficient to accurately account for wave absorption. The ability of the iterative method to find a solution was based on the successful convergence after $100 / \lambda$  iteration steps and the ability to meet a tolerance of $10^{-4}$ with the definition of 'tolerance' as in defined previously in equation \ref{eq:tolerance}. Furthermore analysis of the optical depth can be restricted to purely perpendicular propagating waves as oblique waves are generally absorbed by the plasma over shorter magnetic gradient lengths than pure perpendicular waves.

It has been shown that for low plasma densities (typically $X < 0.25$) no controller is necessary to stabilize the solution. For $X$ in the vicinity of, and in between the branch-points (where FX2-EBW coupling occurs) the original method fails to trace the entirety of the lfs-FX branch. Evaluation of $Y_i$ at different plasma temperatures indicates that relativistic broadening applies to these locations, hence the corresponding $\tilde\tau_i$ scales linearly with temperature, justifying it's normalization based on $T_e$. In Table \ref{Tab:TauNorm} these normalized optical densities are depicted for the previously defined locations of $Y$ at different $X$. Since the original iterative method with $\lambda = 1$ fails close, or prior to $Y = 0.5$, $\tilde\tau_a$ is (close to) zero and wave absorption is not accounted for correctly. After failing of the iterative method, solutions often converge towards the hfs-branch which - having a similar imaginary part - still results in reasonable absorption profiles. It could however occur that solutions of the O-mode are found resulting in a far lower optical density at the X2 resonance layer. A reduction in step-size $\lambda$ allows for a correct evaluation of the branch until $Y_b$ leading to values of $\tilde\tau_b > 1$. Using the full controller allows for the complete evaluation of the branch leading to $\tilde\tau_c > 2$.

\begin{table}[]
\centering
\caption{Normalized optical depth $\tilde\tau$ at locations $Y_{1,2,3}$ for different plasma densities and pure parallel propagation. 'R-cutoff' indicates successful convergence until the R-cutoff.}
\begin{tabular}{llll}
$X$ & $\tilde\tau_1$ & $\tilde\tau_2$ & $\tilde\tau_3$    \\ \hline
$0.3$  & $0.03$   & $1.6$     & $3.18$       \\
$0.4$  & $0$    & $1.22$  & $2.17$ \\
$0.5$  &  $0$  &  R-cutoff & R-cutoff
\end{tabular}
\label{Tab:TauNorm}
\end{table}

Based on the typical normalized optical densities obtained, recommendations can be made regarding the use of a controller based on the product $L_B \; B_0 \; T_e$ [m T keV]. First of all, the original method should be avoided since convergence might fail prior to entering the X2 resonant area. Secondly, a reduction in step-size might be sufficient for larger tokamaks where $L_B \; B_0 \; T_e > 2$, as most of the wave deposition ($> 90\%$) will be accounted for correctly. Finally, the fully controlled method is generally able to account for absorption correctly. Only in the vicinity of points where the FX-mode and the EBW-mode coincide the method might break down. For $L_B \; B_0 \; T_e > 1$ this will however not be an issue and increasing numerical precision through $dY$ or $\lambda$ is able to resolve this. Lastly it should be noted that failure of the iterative method usually results in an increased calculation time, as the routine will continue searching for a solution until a hard limit on the amount of iterations is reached.

\section{Summary \& Conclusions} \label{sec:Summary}
To find solutions of the WPDR it can be cast into a bi-quadratic equation implicit of form on $N_\perp^2$, where a method is used to iteratively search for a solution \cite{Westerhof-1989,Matsuda-1991,Farina-2008}. At the \snd harmonic electron cyclotron resonance the fast extraordinary (FX) mode couples to electron Bernstein waves (EBWs) at higher densities~\cite{ImreWeitzner-1985a,ImreWeitzner-1985b,Egedal-1994}. Treating the iterative method for solving the bi-quadratic equation as a dynamical system, indicates that EBW-solutions can become unstable. FX-mode solution branches that couple to the EBW-mode can therefore not be found using the existing iterative method, which instead converges to undesired solutions.

By means of a reduction of the step-size, and a rotation of the step-direction in the complex plane of $N_\perp^2$ unstable solutions are rendered stable, while increasing overall robustness of the iterative method. Such rotation is implemented in the form of a feed-forward controller. This adjustment allows the accurate determination of points of coinciding FX-mode with EBWs, allowing the mapping of the warm plasma dispersion relation.

The Warm plasma dispersion relation (WPDR) has complicated solution structures around the \snd harmonic cyclotron resonance, that have now been completely mapped by including the analysis for oblique waves. Part of this complicated structure arises from the coupling between the FX mode and EBWs within the \snd harmonic EC resonance layer. By treating solutions to the warm dispersion function as Riemann surfaces in the CMA-plane ($X = \frac{\omega_{pe}^2}{\omega^2}$, $Y = \frac{\omega_{ce}}{\omega}$), points of coinciding FX-mode and EBWs could be mapped on (X,Y) as a function of the electron temperature $T_e$ and the parallel refractive index $N_\parallel$ \cite{Egedal-1994}. These points of coinciding modes are identified as branch points. At the \snd harmonic cyclotron resonance the branch points $P_1$ and $P_2$ encompass the region where the FX-mode is coupled to the EBW-mode. In that region the crossing of the FX-mode and EBW branch is avoided. Instead, X-mode waves originating from the low-field side transform inside the \snd harmonic resonance layer to the high-field side EBW branch and vice versa for X-mode wave originating from the high-field side. For oblique waves this coupling vanishes for absolute values of the parallel wave component $N_\parallel$ above a critical value, as the branch points $P_1$ and $P_2$ merge together. The location of the merging of the branch points roughly scales with $\sqrt[3]{T_e}$ like other relativistic effects.

Evaluation of the optical depth up to a point where the uncontrolled iterative method becomes unstable, shows that stabilization of the iterations is essential for the correct evaluation of extraordinary mode ECRH at the \snd harmonic EC resonance. For large tokamaks ($R_0 \; B_0 \; T_e > 2$ [m T keV] ) a reduction in step-size will suffice as most wave energy is deposited prior to the solution becoming unstable. Outside the \snd harmonic EC resonance layer no controller is required to find solutions to the WPDR.

\section*{Data Availability}
The data underlying this article are available at \href{http://dx.doi.org/10.5281/zenodo.4095258}{http://dx.doi.org/10.5281/zenodo.4095258}.

\begin{acknowledgments}
DIFFER is part of the institutes organisation of NWO. This work has been carried out within the framework of the EUROfusion Consortium and has received funding from the Euratom research and training programme 2014-2018 and 2019-2020 under grant agreement No 633053. The views and opinions expressed herein do not necessarily reflect those of the European Commission.
\end{acknowledgments}

\appendix
\section{}

As mentioned in the main text, the warm plasma dispersion relation can be cast into bi-quadratic form:
\begin{equation}
    A N_\perp^4 - B N_\perp^2 + C = 0,
\end{equation}
where in the general case the coefficients $A$, $B$ and $C$ are themselves dependent on $\nps{}$. Only in case of the fundamental harmonic electron cyclotron resonance and including only the lowest significant order in the finite Larmor radius expansion for the Bessel functions occurring in the dielectric tensor, $\varepsilon$, a pure biquadratic equation formulation can be obtained with the following choice for the coefficients as originally proposed by Fidone et al.~\cite{fidone_1978,Westerhof-1989}:
\begin{eqnarray}
    A \equiv &(\chi_{xz} + N_\parallel)^2 - (\varepsilon_{xx} - N_\parallel^2)(1 - \chi_{zz}) \nonumber \\
    B \equiv &(\varepsilon_{yy} - N_\parallel^2)(\chi_{xz} + N_\parallel)^2 \nonumber \\
             &+ \varepsilon_{xy}^2 (1 - \chi_{zz}) - 2 \varepsilon_{xy} \chi_{yz} (\chi_{xz} + N_\parallel)  \label{biq_coef}\\
             & - (\varepsilon_{xx} - N_\parallel^2) \bigl( \chi_{yz}^2 - \varepsilon_{zz}^0 - (\varepsilon_{yy} - N_\parallel^2) (1 - \chi_{zz}) \bigr) \nonumber \\
    C \equiv & \varepsilon_{zz}(0) \biggl( \varepsilon_{xy}^2 + (\varepsilon_{xx} - N_\parallel^2) (\varepsilon_{yy} - N_\parallel^2) \biggr). \nonumber
\end{eqnarray}
where for $\chi$ we have the definitions
\begin{equation}
    \chi_{xz} \equiv \frac{\epsilon_{xz}}{N_\perp} \mathrm{,} \hspace{0.5cm} \chi_{yz} \equiv \frac{\epsilon_{yz}}{N_\perp} \hspace{0.5cm} \mathrm{and} \hspace{0.5cm} \chi_{zz} \equiv \frac{\epsilon_{zz} - \epsilon_{zz}^o}{N_\perp^2} \mathrm{,}
\end{equation}
and
\begin{equation}
    \epsilon_{zz}^o \equiv 1 - \frac{\omega_p^2}{\omega} \mathrm{.}
\end{equation}
For the second and higher harmonics or taking into account more than one term in the Bessel function expansions, the coefficients in this biquadratic formulation become themselves dependent on $\nps{}$ resulting, in principle, in a higher order equation introducing additional solutions: the Bernstein modes. The biquadratic formulation, however remains a useful tool allowing to solve the dispersion relation iteratively as shown in the present paper.
Note that in the notation of the dielectric tensor elements the usual convention is followed with the $z$ axis aligned to the equilibrium magnetic field and the $x$ axis aligned to the perpendicular component of the wave vector.

The two solutions of the biquadratic equation are
\begin{equation}
    \nps = B \pm \frac{\sqrt{B^2-4AC}}{2A},
\end{equation}
where with the present algebraic formulation of the coefficients the $+$ sign refers to the O-mode solution and the $-$ sign to the X-mode solution. Note however that any a numerical implementation should take proper care of selecting the correct Riemann sheet when evaluating the square root function. In particular, when the determinant is seen to cross the negative real axis when tracing a particular solution, the numerical implementation will have to toggle the sign of the evaluated square root function.

\nocite{*}
\bibliography{paper}

\end{document}